\newcommand{\be}{\begin{equation}}
\newcommand{\ee}{\end{equation}}
\newcommand{\bea}{\begin{eqnarray}}
\newcommand{\eea}{\end{eqnarray}}
\newcommand{\ba}[1]{\begin{array}{#1}}
\newcommand{\ea}{\end{array}}

\newcommand{\diracslash}[1]{#1\llap{/\kern2pt}}

\documentclass[12pt]{iopart}
\usepackage{iopams}  
\usepackage{graphicx}
\usepackage{epsfig}
\usepackage{amssymb}
\usepackage{times}
\usepackage{amssymb}
\usepackage{mathrsfs}
\usepackage{graphicx}
\usepackage{epsfig}
\usepackage{dcolumn}
\usepackage{color}
\usepackage{bm}
\begin{document}
\setlength{\topmargin}{0.2in}

\title{Interactions and low energy collisions between an alkali ion and an alkali atom of different nucleus}
\author{Arpita Rakshit$^{1}$, Chedli Ghanmi$^{2,3}$, Hamid Berriche$^{2,4}$,  and Bimalendu Deb$^{5}$}
\address{$^1$ Sidhu Kanhu Birsa Polytechnic,
Keshiary, Paschim Medinipur 721133, India.\\
$^2$ Laboratory of Interfaces and Advanced
Materials, Physics department, Faculty of Science, University of
Monastir, 5019 Monastir, Tunisia.\\
$^3$ Physics Department, Faculty of Science,
King Khalid University, P. O. Box 9004, Abha, Saudi Arabia.\\
$^4$Mathematics and Natural Sciences Department, School of Arts and Sciences,
American University of Ras Al Khaimah, Ras Al Khaimah,  P. O. Box 10021, RAK, UAE\\
$^5$Department of Materials Science, Raman Center for Atomic, Molecular and
Optical Sciences, Indian Association for the Cultivation of Science,
Jadavpur, Kolkata 700032, India. }
\begin{abstract}
We study theoretically interaction potentials and low energy
collisions between different alkali atoms and alkali ions.
Specifically, we consider systems like X + Y$^{+}$, where X(Y$^{+})$
is either Li(Cs$^+$) or Cs(Li$^+$), Na(Cs$^+$) or Cs(Na$^+$) and
Li(Rb$^+$) or Rb(Li$^+$). We calculate the
molecular potentials of the ground and first two excited states of
these three systems using pseudopotential method and compare our
results with those obtained by others. We derive  ground-state
scattering wave functions and analyze cold collisional properties  of these systems for a
wide range of energies. We find that, in order to get  convergent
results for  the total scattering cross sections for  energies of
the order 1 K, one needs to take into account at least 60 partial
waves. Low energy scattering properties calculated in this paper may serve 
as a precursor for experimental exploration of quantum collisions between an 
alkali atom and an alkali ion of different nucleus. 
\end{abstract}

\maketitle

\section{Introduction}

Over more than last three decades, there has been tremendous progress in the technology of cooling and trapping of single-electron 
alkali atomic gases and 
some two-electron atomic gases.  In parallel, the developments in the technology of trapping and cooling of some ions, 
specially  alkaline-earth cations over last several decades has led to new vistas of research activities with trapped and cold ions. 
With recent experimental developments 
\cite{prl:2009:vuletic,aymar:prl:2011,molphys,hudson:prl:2011,hudson:prl:2012,Zipkes1,Zipkes2,Zipkes3,Mukaiyama:pra:2013,schmid:2012:rsi,Denschlag:prl:2012,Narducci:pra:2012,schmid,rangwala:apl:2012,rangwala:natcom:2012,rangwala:pra:2013} 
using hybrid ion-atom traps
\cite{hytrap},  both atoms
and ions can be confined simultaneously in a common space or an atom-trap can be 
merged with an ion-trap. These hybrid systems open up  prospects 
for emulating  solid-state physics with laser-generated periodic structure of ions \cite{prl111:2013:schmidt-kaler} and 
for exploring physics of ion-controlled Josephson junction \cite{prl109:2012:schmidt-kaler,pra89:2014:negretti}, 
Tonk-Girardeau gas with 
an ionic density ``bubble'' \cite{pra81:2010:busch}.

At a fundamental level, these new hybrid devices 
facilitate the laboratory investigations on ion-atom collisions
in hitherto unexplored low energy regimes -- from milliKelvin down to microKelvin or sub-microKelvin temperature regime.      
Ion-atom scattering at 
low energy is important for a number of physical systems, such as
cold plasma, planetary atmospheres, interstellar clouds.
Gaining insight into the ion-atom interactions and scattering at
ultra-low energy down to Wigner threshold law regime is important to
understand charge transport\cite{cote2000} at low
temperature, radiative association \cite{aymar:prl:2011,molphys,njp17:2015:aymar}, ion-atom bound states\cite{bound}, 
ion-atom photoassociation\cite{Rakshit2011,jcp:2011:dulieu}, and many other related phenomena.
Several theoretical investigations  
\cite{arXiv:1409.1192,pra91goodman,njp,Zhang2009,Zygelman1989,pra79,bgao,bgao2012} of atom-ion cold collisions have
been carried out in recent times. It is proposed that controlled
ion-atom cold collision can be utilized for quantum information
processes\cite{pra81:2010:calarco,contphys}. Several recent theoretical 
studies  have 
focused on charge-transfer \cite{pccp13:2011:belyaev,pra87:2013:belyaev,jpb47:2014:McCann} and 
chemical reaction processes \cite{jpb47:2014:zygelman,pra86:2012:McCann} at ultralow energies between an alkaline-earth 
ion or Yb$^+$ and an alkali atom.

Ions are usually trapped by radio-frequency fields. The major
hindrance to cool trapped ions below milliKelvin temperature stems
from the trap-induced micro-motion of the ions, which seems to be
indispensable for such ion traps. Therefore, it is difficult to
achieve sub-microKelvin temperature for an ion-atom system in a
hybrid trap. But, to explore fully quantum or Wigner
threshold law regime for ion-atom collisions, it is essential to
reduce the temperature of  ion-atom hybrid systems below one
micro-Kelvin. One way to overcome this difficulty is to device new
methods to trap and cool both ions and atoms in an optical trap.
With recent experimental demonstration of an optical trap for ions\cite{naturecomm}, the prospect for experimental explorations of
ultracold ion-atom collisions in Wigner threshold regime  appears to
be promising.

In the contexts of ion-atom cold collisions of current interest,
there are mainly four types of ion-atom systems:(1) an alkaline-earth ion  interacting with an alkali metal atom, (2) a rare earth
ion (such as Yb) interacting with an alkali metal atom, (3) an
alkali metal ion interacting with an alkali atom of the same nucleus
and (4) an alkali metal ion interacting with an alkali atom of
different nucleus. In the first two types, both the ion and the atom
have one valence electron in the outermost shell. In the third type,
the ion has closed shell structure while the atom has one valence
electron in the outermost shell. Since both the ion and atom are of
the same nucleus, there is center of symmetry for the electronic
wave function of the ion-atom pair. Because of this symmetry,
resonant charge transfer collision is possible in the third type. In
the fourth type, the ion is of closed shell structure while the atom
has one valence electron in the outermost shell. Since the nuclei of
the atom and the ion are different, there is no resonant charge
transfer collision in fourth type, though non-resonant charge
transfer is possible in all the types. For scattering in the
ground-state potentials, non-resonant charge transfer
collision is suppressed  in the ultralow energy regime ($<\mu$K). \\

\begin{figure}
        \includegraphics[width=10cm]{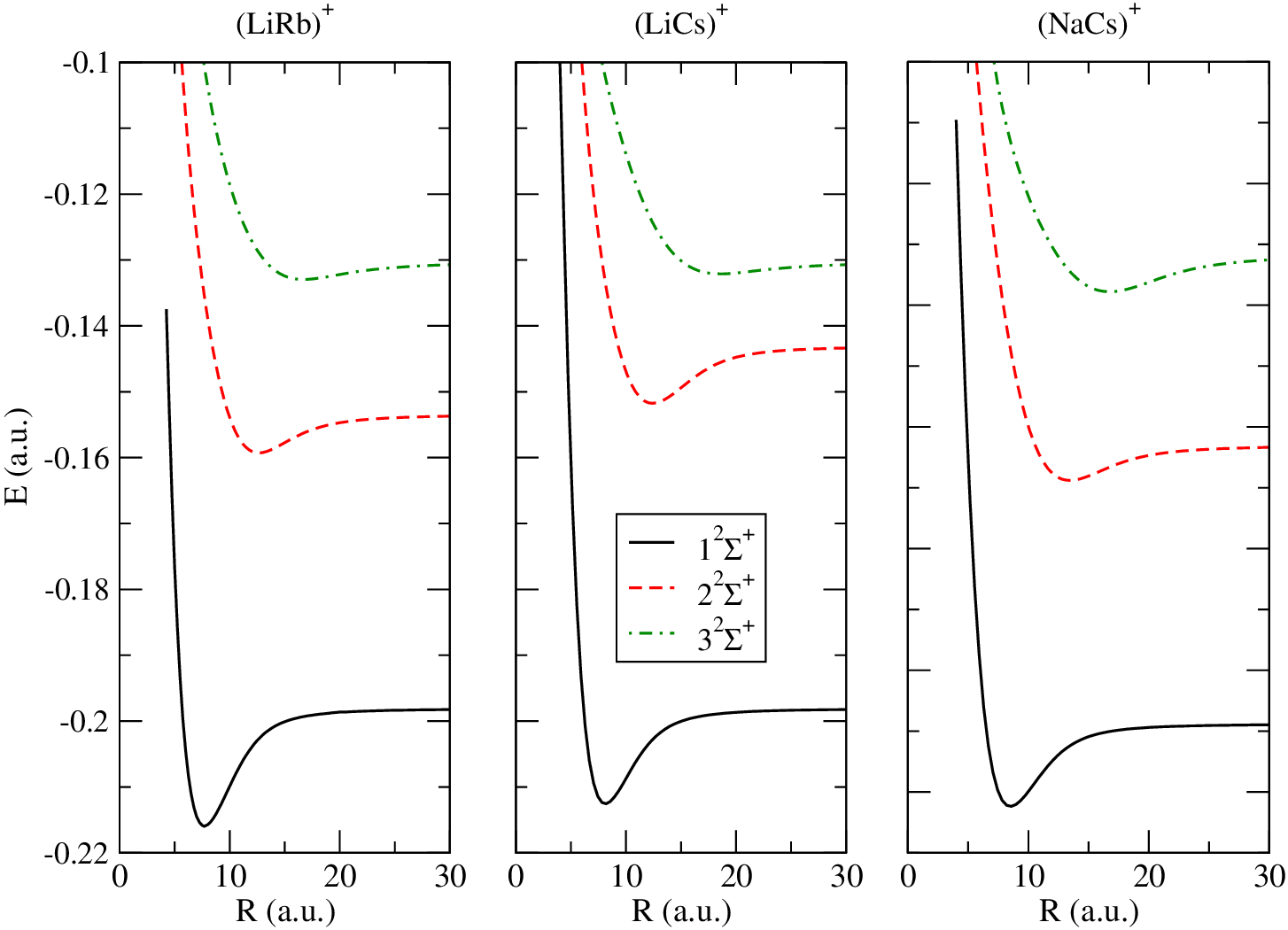}
        \caption{Adiabatic potential curves for (LiRb)$^{+}$, (LiCs)$^{+}$ and (NaCs)$^{+}$}
\end{figure}

Here we study cold collision in the fourth type of ion-atom systems, though 
these systems are not straightforward to obtain
experimentally. Normally,  neutral alkali-metal atoms and alkaline-earth ions can be
laser-cooled and trapped. However, there are some recent experimental studies showing 
different techniques \cite{schmid:2012:rsi,Denschlag:prl:2012,rangwala:apl:2012,rangwala:natcom:2012} 
for trapping and cooling of closed-shell akali-metal ions. 
Our primary aim is to understand ground-state elastic scattering
processes between  such an ion and an alkali atom of different nuclei at ultralow energy where
inelastic charge transfer process can be ignored. A proper understanding of ground-state
elastic scattering processes\cite{bgao,bgao2012} at low energy in
an ion-atom system is important for exploring coherent control of
ion-atom systems and formation of ultracold molecular ions by
radiative processes. Due to the absence
of resonance charge transfer reaction, the fourth type system is
preferable for our purpose. However, there remains finite possibility for inelastic collisions if the
hyperfine structure of the neutral atom is taken in account. In our calculations we 
do not consider hyperfine interactions. 

Let a cold alkali metal ion {\bf A}$^{+}$ interact with a cold
alkali metal atom {\bf B}. Suppose, the atomic mass of {\bf B} is
smaller than that of atom {\bf A}. Then, for {\bf AB}$^{+}$
molecular system, ground-state continuum asymptotically corresponds
to the atom {\bf B} (ns) in $^{2}$S electronic state and the ion
{\bf A}$^{+}$ (complete shell) in $^{1}$S electronic state. In the
separated atom limit, the first excited molecular potential
asymptotically goes as the atom {\bf A} (ns) in $^{2}$S electronic
state and alkali ion {\bf B}$^{+}$ (complete shell) in $^{1}$S
electronic state while the second electronic excited potential
corresponds to {\bf B} (np) in $^{2}$P electronic state and alkali
ion {\bf A}$^{+}$ (complete shell) in $^{1}$S electronic state.

\begin{table}[ph]   
\caption{A comparison of the spectroscopic constants for the ground (X$^{2}\Sigma^{+}$) and the first and second
     excited (2$^{2}\Sigma^{+}$ and 3$^{2}\Sigma^{+}$) electronic states of (LiRb)$^{+}$ molecular ion with the available works.}
{\begin{tabular}{@{}llllllll@{}} \hline 
\\[-1.8ex] 
State &  $R_{e}$(a.u.) &$D_{e}$(cm$^{-1}$) & T$_{e}$(cm$^{-1}$)&$\omega_{e}$(cm$^{-1}$) & $\omega_{e}\chi_{e}$(cm$^{-1}$)&$B_{e}$(cm$^{-1}$)&{references}\\[0.8ex] 
\hline \\[-1.8ex] 
X$^{2}\Sigma^{+}$ & 7.70 & 3912 &  & 137.56 & 1.67 & 0.158098  & This work  \\
              &      & 3705 &  &        &      &           & \cite{Bellomonte1974}\\
              & 7.50 &      &  &        &      &           & \cite{Carlson1980}\\
              & 7.60 & 3432 &  &        &      &           & \cite{Patil2000} \\
              & 7.54 & 4193 &  & 139.65 &      &           & \cite{Azizi1}\\
2$^{2}\Sigma^{+}$ & 12.67 & 1270 &  12438 & 63.73 &  0.51    &  0.058456 & this work\\
              & 12.60 & 1306 &        & 63.00 &          &           & \cite{Azizi1} \\
3$^{2}\Sigma^{+}$ & 16.66 & 617 & 18200& 35.73 & 1.55 & 0.033769 &This work \\[0.8ex] 
\hline \\[-1.8ex] 
\multicolumn{8}{@{}l}{ }\\
\end{tabular}}
\label{tab1}
\end{table}

We investigate cold collisions over a wide range of energies  in
three different alkali atom-alkali ion systems in the ground
molecular potentials. The colliding atom-ion pairs we consider are:
(i) Cs$^{+}$ + Li, (ii) Cs$^{+}$ + Na, and (iii) Rb$^{+}$ + Li.
Since at ultralow energy, charge transfer reaction is highly
suppressed in these systems, we study elastic scattering only. 
 To
calculate scattering wave functions, we need the data for
Born-Oppenheimer adiabatic potentials of the systems. In  the long
range where the separation r $>$ 20 $a_{0}$ ($a_{0}$ is the Bohr
radius), the potential is given by the sum of the dispersion terms,
which in the leading order goes as $1/r^{4}$. We obtain short-range
potentials by pseudopotential method. The short-range part is
smoothly combined with the long-range part to obtain the potential
for the entire range. We then solve time-independent Schroedinger
equation for these potentials with scattering boundary conditions by
Numerov-Cooley \cite{cooley1961} algorithm. We present detailed results of scattering
cross sections for these three systems for
energies ranging 
from 0.01 micro-Kelvin ($\mu$K) to 1 Kelvin (K). However, the low energy regime of our interest ranges from 
sub-$\mu$K to 1 milli-Kelvin (mK). Here we choose to use Kelvin as the unit of energy to readily 
convey the information how cold the system should be to explore such collision physics. Normally 
the results of atomic collisions are expressed in atomic unit (a.u.). 
Note that 1 a.u. of energy corresponds to  3.1577465  $\times$ 10$^5$ K or 
27.21138386 electron-Volt (ev). \\ 

\vspace{0.5cm}

\begin{figure}
        \includegraphics[width=3.65in]{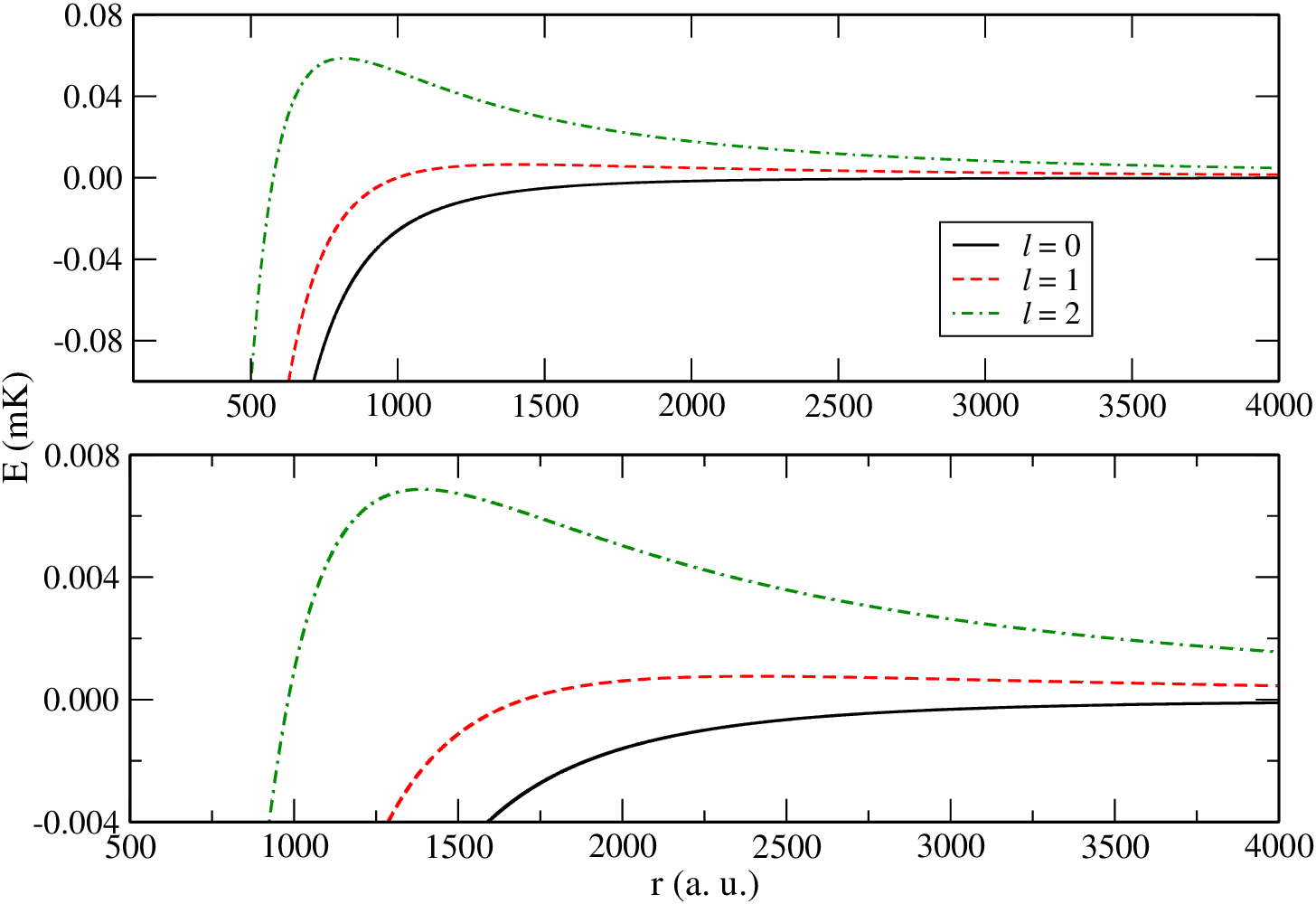}
        \caption{(Color online) In the upper panel,  the centrifugal
energies in unit of milliKelvin (mK) for $s$- (black, solid), $p$- (red, dashed) and $d$- (green,
dashed-dotted) partial waves  are plotted against atom-ion separation
for $1^{2}\Sigma^{+}$ state of (LiCs)$^{+}$. The lower panel
shows the same for (NaCs)$^{+}$.}
\end{figure} 

\section{Interactions and molecular properties}

Under Born-Oppenheimer approximation, we compute the adiabatic
potential energy curves of the $1^{2}\Sigma^{+}$, $2^{2}\Sigma^{+}$
and $3^{2}\Sigma^{+}$ electronic states of the three ionic molecules
(LiRb)$^{+}$, (LiCs)$^{+}$ and (NaCs)$^{+}$ using the
pseudopotential method proposed by Barthelat and Durand\cite{Barthelat1975} 
in its semi-local form and used in many previous works\cite{Ghanmi1,Ghanmi2,Ghanmi2012,Berriche2003,Berriche2005}. The
interaction potentials between different alkali metal ion and an
alkali metal atom were calculated by Valance\cite{Valance1978} in
1978 by the method of pseudopotential. 
Here we briefly describe spectroscopic constants and interaction  potentials of the three considered atom-ion systems, 
and the pseudopotential method  used to obtain these molecular properties.   Our final goal is to calculate 
cold collisional properties of these systems using these potentials as described in the next section. \\

\vspace{0.5cm}

\begin{table}[ph]   
\caption{A comparison of the spectroscopic constants for the ground (X$^{2}\Sigma^{+}$) and the first and second
     excited (2$^{2}\Sigma^{+}$ and 3$^{2}\Sigma^{+}$) electronic states of (LiCs)$^{+}$ molecular ion with the work of
     Khelifi et al.$^{57}$}
{\begin{tabular}{@{}llllllll@{}} \hline 
\\[-1.8ex] 
State &  $R_{e}$(a.u.) &$D_{e}$(cm$^{-1}$) & T$_{e}$(cm$^{-1}$)&$\omega_{e}$(cm$^{-1}$) & $\omega_{e}\chi_{e}$(cm$^{-1}$)&$B_{e}$(cm$^{-1}$)&{references}\\[0.8ex] 
\hline \\[-1.8ex] 
1$^{2}\Sigma^{+}$ & 8.12 & 3176 & 0 & 124.46 & 1.08 & 0.138337 & This work   \\
              & 7.19 & 3543 &  &  &  &  & \cite{Khelifi2011}  \\
2$^{2}\Sigma^{+}$ & 12.42 & 1911 &  13343 & 68.69 &  0.63 &  0.059126 & This work \\
             & 12.37 & 2022 & 19553 &  &  &  & \cite{Khelifi2011}  \\
3$^{2}\Sigma^{+}$ & 18.53 & 422 & 17660 & 28.07 & 1.29 & 0.026570 & This work  \\
              & 18.38 & 409 & 20849 &  &  &  & \cite{Khelifi2011}  \\
\hline \\[-1.8ex] 
\multicolumn{8}{@{}l}{ }\\
\end{tabular}}
\label{tab2}
\end{table}

\subsection{Adiabatic Potentials}

For the three ionic systems (LiRb)$^{+}$, (LiCs)$^{+}$ and
(NaCs)$^{+}$, the potential energy curves of the 1-3$^{2}\Sigma^{+}$
electronic states, are built from an ab initio calculation for the
internuclear distances ranging from 3 to 200 $a_0$ ($a_0$ is Bohr radius)  as discussed in the  sub-section 2.2. 
These states dissociate
respectively, into Li(2s and 2p) + Rb$^{+}$ and Rb(5s) + Li$^{+}$,
Li(2s and 2p) + Cs$^{+}$ and Cs(6s) + Li$^{+}$, and Na(3s and 3p) +
Cs$^{+}$ and Cs(6s) + Na$^{+}$. They are displayed in figure 1(a)
for LiRb$^{+}$, 1(b) LiCs$^{+}$ and 1(c) for NaCs$^{+}$. 
In the long range part 
beyond $ 200 a_0$,
the potential is given by the sum of the dispersion terms. The short
range part ($\le 20 a_0$) of ab initio potential is smoothly combined with the long-range part by interpolation 
to obtain the potential for entire range extending to several thousand $a_0$.
For the
three ionic systems, we remark, that the ground state has the
deepest well compared to the 2$^{2}\Sigma^{+}$ and 3$^{2}\Sigma^{+}$
excited states. Their dissociation energies are of the order of
several 1000 cm$^{-1}$. Their equilibrium positions lie at
separations that are relatively larger than those of typical neutral
alkali-alkali diatomic molecules. \\ \\

\begin{figure}
\includegraphics[width=3.65in]{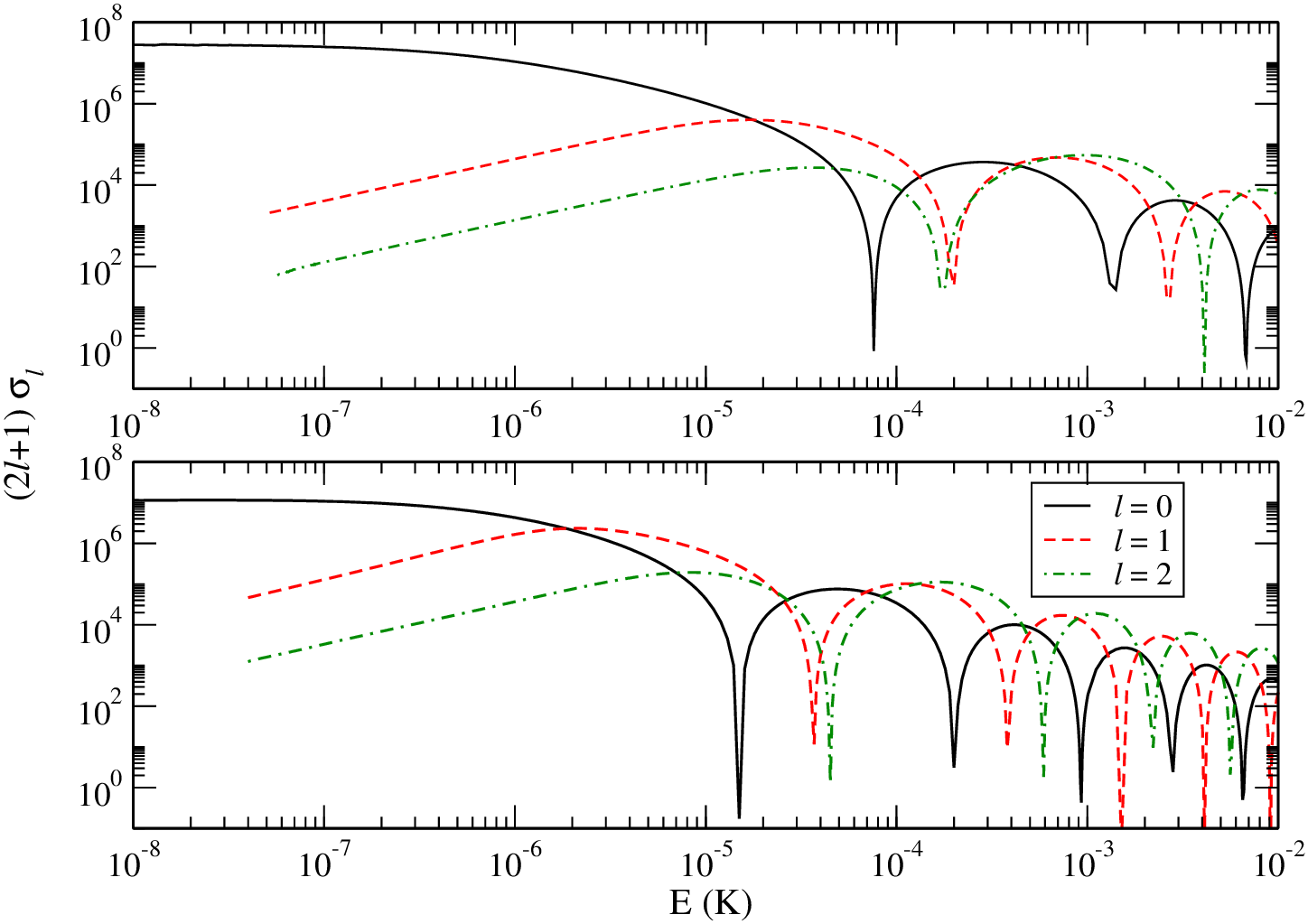}
\caption{(Color online) $s$-
$p$- and $d$-wave  scattering cross sections in atomic unit (a.u.) or in unit of $a_0^2$ ($a_0$ is Bohr radius) are plotted against
collision energy $E$ in Kelvin (K) for $1^2\Sigma^+$ state of (LiCs)$^+$  and  (NaCs)$^+$ in upper and lower panel, respectively.}
\end{figure}

The long range potential is given by the expression
\begin{equation}
V(r) = -\frac{1}{2}\left( \frac{C_{4}}{r^{4}} + \frac{C_{6}}{r^{6}}
+ \cdots \right)
\end{equation}
where $C_{4}$, $C_{6}$ correspond to dipole, quadrupole
polarisabilities of concerned atom. Hence, the long range
interaction is predominately governed by polarisation interaction.
Dipole polarisabilities for Na (3s) and Li (2s) are 162 a.u. and
164.14 a.u., respectively. One can define a  characteristic length
scale of the long-range  potentials by $\beta = \sqrt{2\mu
C_{4}/\hbar^{2}}$. The values of $\beta$ for the collision of Li -
Cs$^{+}$ pair and Na - Cs$^{+}$ pair are 1411.5 a.u. and 2405.8
a.u., respectively

For the ground state 1$^{2}\Sigma^{+}$ the well depth and the
equilibrium position are 2979 cm$^{-1}$ and 8.51 a.u., respectively,
for (NaCs)$^{+}$ while those for (LiCs)$^{+}$ are 3176 cm$^{-1}$ and
8.12 a.u., respectively. Reduced mass for (NaCs)$^{+}$ and
(LiCs)$^{+}$ (for $^7$Li isotope)  are taken as 19.5995 a.u. and 6.6642 a.u.,
respectively. For both $^{6}$Li-Rb$^{+}$ and $^{6}$Li-Rb$^{+}$
isotopes, we used the reduced masses 5.6171 and 6.4805 a.u.
respectively.

\subsection{Pseudopotential method: Results and discussions}

The use of the pseudopotential method, for each (XY)$^{+}$ ionic
molecules (LiRb$^{+}$, LiCs$^{+}$ and NaCs$^{+}$), reduce the number
of active electrons to only one electron. We have used a core
polarization potential $V_{CPP}$ for the simulation of the
interaction between the polarizable X$^{+}$ and Y$^{+}$ cores with
the valence electron. This core polarization potential is used
according to the formulation of M\"{u}ller et al.\cite{Muller1984},
and is given by
\begin{equation}
V_{CPP} = -\frac{1}{2} \Sigma_{\lambda}
\alpha_{\lambda}\vec{f}_{\lambda} \cdot \vec{f}_{\lambda}
\end{equation}
where $\alpha_{\lambda}$ and $\vec{f}_{\lambda}$  are respectively
the dipole polarizability of the core $\lambda$, and the electric
field produced by valence electrons and all other cores on the core
$\lambda$. The electric field $\vec{f}_{\lambda}$ is defined as:
\begin{equation}
\vec{f}_{\lambda} = \Sigma_{i} \frac{\vec{r}_{i\lambda}}{r^{3}_{i}} F(\vec{r}_{i\lambda}, \rho_{\lambda})-\Sigma_{\lambda'
\neq \lambda} \frac{\vec{R}_{\lambda', \lambda}}{R^{3}_{\lambda', \lambda}}Z_{\lambda}
\end{equation}
where $\vec{r}_{i\lambda}$ and $\vec{R}_{\lambda', \lambda}$ are
respectively the core-electron vector and the core-core vector. \\

\vspace{0.5cm}

\begin{table}[ph]   
\caption{A comparison of the spectroscopic constants for the ground (X$^{2}\Sigma^{+}$) and the first and second
     excited (2$^{2}\Sigma^{+}$ and 3$^{2}\Sigma^{+}$) electronic states of (NaCs)$^{+}$ molecular ion with the work
     of Valance$^{49}$.}
{\begin{tabular}{@{}llllllll@{}} \hline 
\\[-1.8ex] 
State &  $R_{e}$(a.u.) &$D_{e}$(cm$^{-1}$) & T$_{e}$(cm$^{-1}$)&$\omega_{e}$(cm$^{-1}$) & $\omega_{e}\chi_{e}$(cm$^{-1}$)&$B_{e}$(cm$^{-1}$)&{references}\\[0.8ex] 
\hline \\[-1.8ex] 
1$^{2}\Sigma^{+}$ & 8.51 & 2979 & 0 & 68.17 & 0.32 & 0.042418 & This work    \\
              & 7.60 & 3388 &  &  &  &  & \cite{Valance1978}  \\
2$^{2}\Sigma^{+}$ & 13.45 & 1262 &  11758 & 33.98 &  0.30 &  0.016977 & This work \\
              & 14.20 & 774 &  &  &  &  & \cite{Valance1978}  \\
3$^{2}\Sigma^{+}$ & 16.77 & 1393 & 18553 & 26.20 & 1.69 & 0.010920 & This work  \\
              & 15.00 & 732 &  &  &  &  & \cite{Valance1978}  \\
\hline \\[-1.8ex] 
\multicolumn{8}{@{}l}{ }\\
\end{tabular}}
\label{tab3}
\end{table}
\vspace{-1cm}

Based on the formulation of Foucrault et al.\cite{Fouc1992} the
cut-off function F($\vec{r}_{i\lambda}, \rho_{\lambda}$) is a
function of the quantum number l. In this formulation, the
interactions of valence electrons of different spatial symmetry with
core electrons are considered in a different way. The used cut-off
radii of the lowest valence s, p, d and f one-electron for the Li,
Na, Rb and Cs atoms are taken from ref.\cite{Ghanmi1,Ghanmi2012,Berriche2003,Berriche2005}. The extended
Gaussian-type basis sets used for Li, Na, Rb and Cs atoms are
respectively (9s8p5d1f/8s6p3d1f)\cite{Berriche2003}, (7s6p5d3f/6s5p4d2f)\cite{Berriche2003},
(7s4p5d1f/6s4p4d1f)\cite{Pavolini1989} and (7s4p5d1f/6s4p4d1f)\cite{Ghanmi2}. The used core
polarizabilities of Li$^{+}$, Na$^{+}$, Rb$^{+}$ and Cs$^{+}$ cores,
which are equal respectively 0.1915 $a_{0}^{3}$, 0.993 $a_{0}^{3}$,
9.245 $a_{0}^{3}$ and 15.117 $a_{0}^{3}$ are taken from\cite{Muller1984}. 
Using the pseudopotential technique each molecule
is reduced to only one valence electron interacting with two cores.
Within the Born-Oppenheimer approximation an SCF calculation,
provide us with accurate potential energy curves and dipole
functions. \\ 

\vspace{0.5cm}

\begin{figure}
        \includegraphics[width=3.65in]{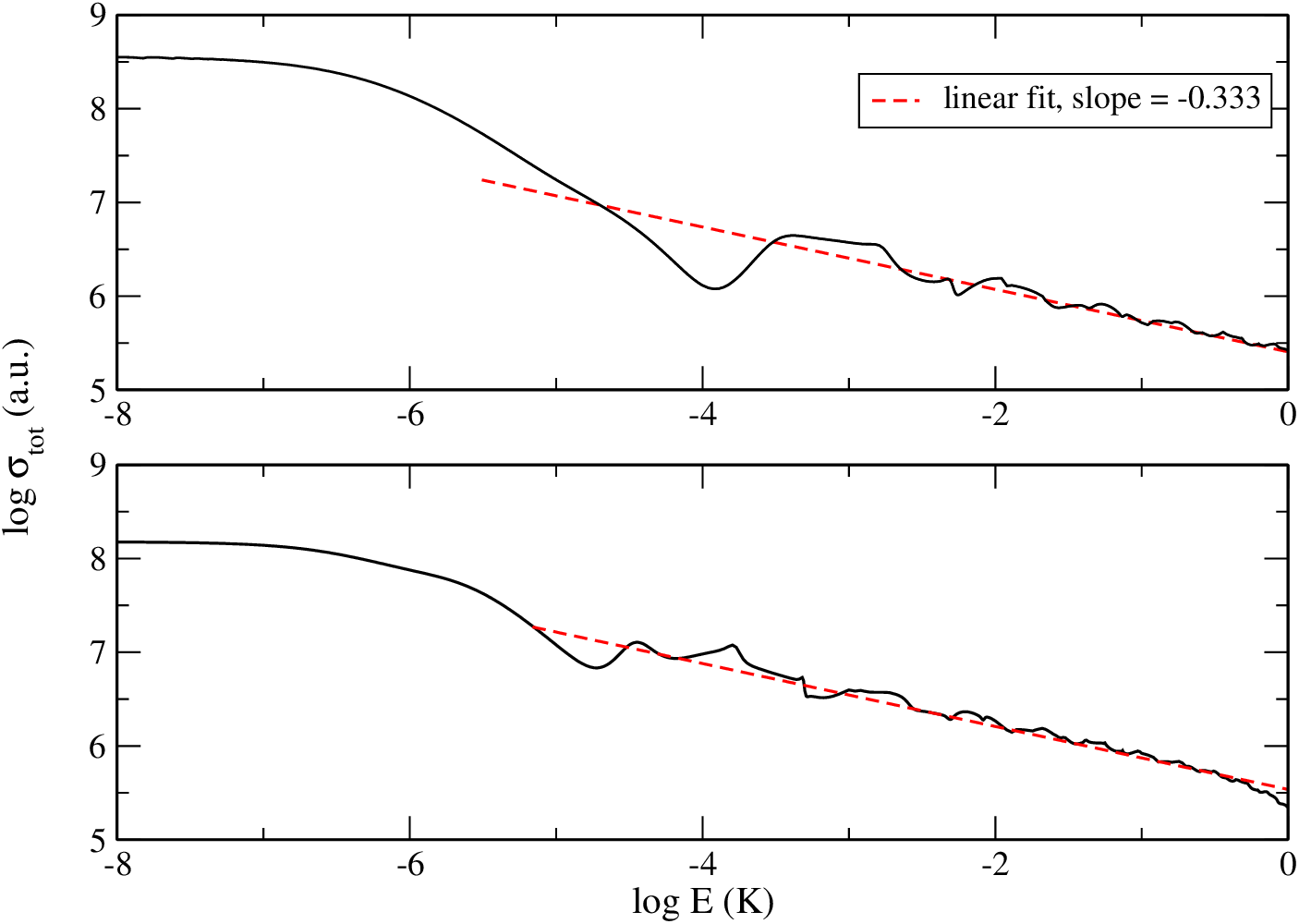}
        \caption{(Color online) Logarithm of total scattering cross section in a.u. as a function of logarithm of 
        energy $E$ in K for electronic
        ground-state 1$^{2}\Sigma^{+}$ of (LiCs)$^{+}$ and (NaCs)$^{+}$ are plotted in upper and lower pannel, respectively.}
\end{figure}

\subsection{Spectroscopic constants}

The spectroscopic constants ($R_{e}$, $D_{e}$, T$_{e}$,
$\omega_{e}$, $\omega_{e}\chi_{e}$, $B_{e})$ of the
1-3$^{2}\Sigma^{+}$ electronic states are presented in tables 1-3
for (LiRb)$^{+}$, (LiCs)$^{+}$ and (NaCs)$^{+}$ respectively. To the
best of our knowledge, no experimental data has been found for these
systems. We compare our spectroscopic constants only with the
available theoretical results. Table 1 presents the spectroscopic
constants of (LiRb)$^{+}$ compared with the other theoretical works\cite{Bellomonte1974,Carlson1980,Patil2000,Azizi1}. As it seems
from table 1, there is a good agreement between the available
theoretical works\cite{Bellomonte1974,Carlson1980,Patil2000,Azizi1} and our ab initio study. Azizi et al.\cite{Azizi1}
reported the spectroscopic constants for many ionic alkali dimers
but only for the ground and the first excited states. They have used
in their study a similar formalism as used in our work. We remark
that our ground state equilibrium distance ($R_{e}$) presents a
satisfying agreement as well as the well depth ($D_{e}$) with the
work of Azizi et al.\cite{Azizi1}. We found for ($R_{e}$) and
($D_{e}$), respectively, 7.70 a. u. and 3912 cm$^{-1}$ and they
found 7.54 a.u. and 4193 cm$^{-1}$. The difference between the results of Azizi {\it et al.} [57] 
and our values are about 2.12\% and 6.70\% for $R_{e}$ and $D_{e}$, respectively.
The same agreement is observed
between our harmonacity frequency ($\omega_{e}$) and that of Azizi
et al.\cite{Azizi1}. The difference between the two values is
2.09 cm$^{-1}$, which represents a difference of 1.49\%. There is also a very good agreement between our
equilibrium distance and that of Patil et al.\cite{Patil2000}
($R_{e}$= 7.60 a.u.), in contrast to their well depth\cite{Patil2000}, which is underestimated ($D_{e}$=3432 cm$^{-1}$).
For the first excited state, Azizi et al.\cite{Azizi1} reported
the spectroscopic constants $R_{e}$=12.60 a.u., $D_{e}$ = 1306
cm$^{-1}$ and $\omega_{e}$ = 63.00 cm$^{-1}$ to be compared with our
values of, respectively, 12.67 a.u., 1270 and 63.73 cm$^{-1}$. The differences in percentages between the values obtained by Azizi {\it et al.} [57] 
and our results are 0.55\%, 2.75\% and 1.14\% for $R_e$, $D_e$ and $\omega_{e}$, respectively.\\ \\

\begin{figure}
        \includegraphics[width=3.65in]{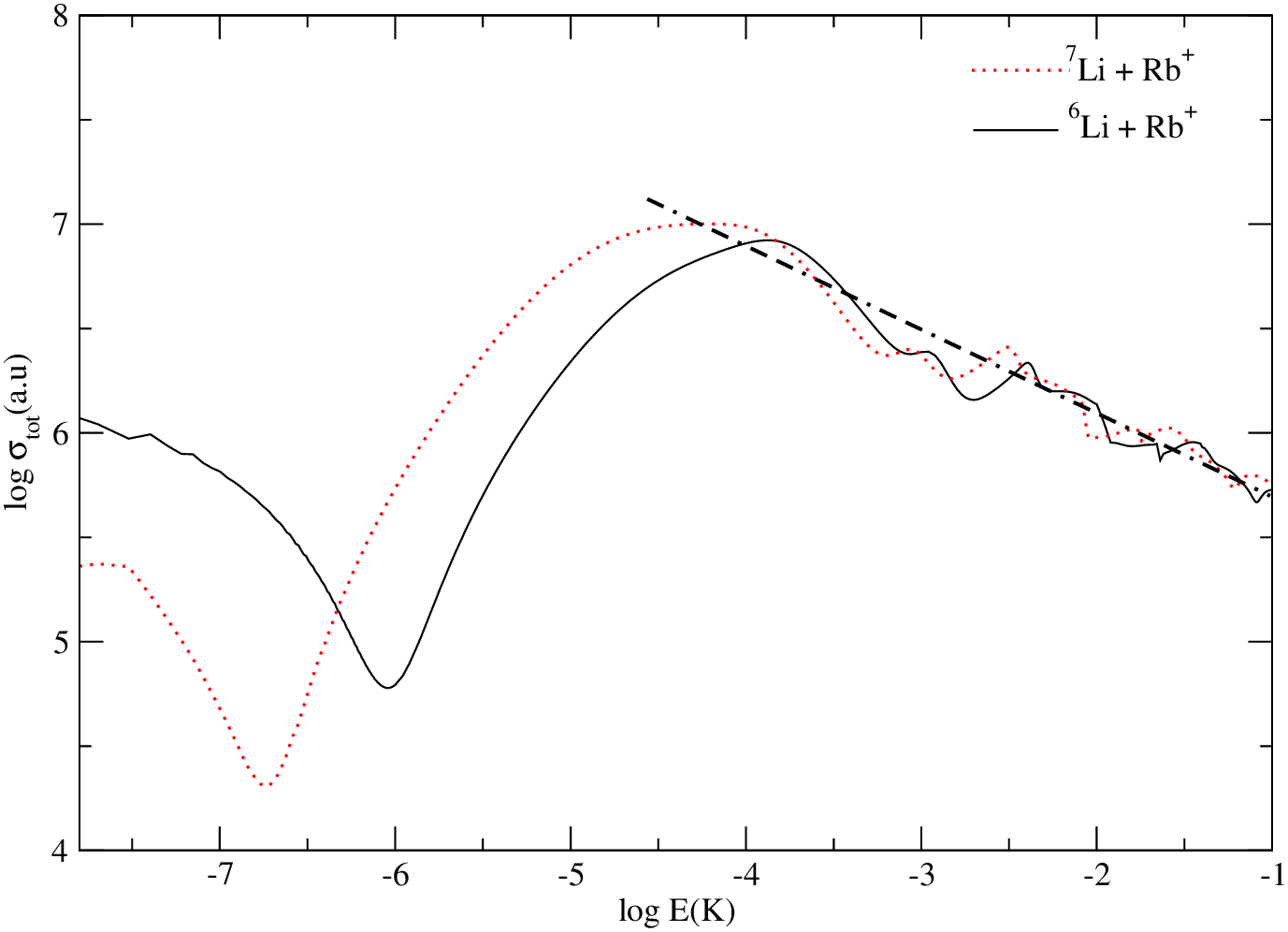}  
        \caption{(Color online) Same as in figure 4 but for  Li + $^{87}$Rb$^{+}$ (1$^{2}\Sigma^{+}$)
        collision. The black solid line and red dotted lines are for $^{6}$Li and  $^{7}$Li,
        respectively. Black dashed lines represent linear fit to both the curves for energies greater 
        than 10$^{-6}$ K.
        Both cases follow one-third law and the intercepts are found to be 3.587 and 3.606 in a.u. 
        for $^{6}$Li and $^{7}$Li, respectively;  where as the
        theoretically calculated values are 3.3269 and 3.3476 in a.u., respectively.}
\end{figure}

The spectroscopic constants of (LiCs)$^+$ and (NaCs)$^+$, are presented respectively in tables 2 and 3 and compared 
with the available theoretical results\cite{Khelifi2011,Valance1978}. To the best of our knowledge there is no experimental data 
for these ionic molecules. For the (LiCs)$^+$ ionic molecule, we compare our results only with the 
theoretical work of Khelifi et al.\cite{Khelifi2011} where they have used a similar method as used in our work. 
They reported for the ground state 1$^2\Sigma^{+}$ the spectroscopic constants $R_e = 7.91$ a.u. and $D_e = 3543$ cm$^{-1}$ to be compared with our values of, respectively, $R_e = 8.12$ a.u. and $D_e = 3176$ cm$^{-1}$.
A rather good agreement is observed for the equilibrium distance, however their potential is much deeper.
Reasonably good agreement between our spectroscopic constant and those of Khelifi {\it et al.} \cite{Khelifi2011} 
is observed for the 2$^2\Sigma^+$ and 3$^2\Sigma^+$ states. The difference between Khelifi {\it et al.} spectroscopic constants and 
our results are found to be  2.58\% and 10.35\% for $R_e$ and $D_e$,  respectively.
For the ground state of (NaCs)$^+$ ionic molecule, there is a good agreement between our well-depth, 
as well as the equilibrium distance with the theoretical results of Valance\cite{Valance1978}.
We found a well-depth of 2979 cm$^{-1}$ located at 8.51 a u., while valance\cite{Valance1978} found 3388 cm$^{-1}$ located at 7.60 a.u.. 
The difference in percentages between valance’s data and our values are 12.07\% and 11.97\% for $R_e$ and $D_e$, respectively.
In contrast to the ground state, 
where the agreement between our work and those of Valance\cite{Valance1978} is good, 
there is a disagreement for the 2$^2\Sigma^+$ and 3$^2\Sigma^+$ states. 
The two states exhibit small potential wells in Valance’s work equal, respectively,  to 774 and 732 cm$^{-1}$;
while in our work we have found significantly higher well depths of 1262 and 1393 cm$^{-1}$, respectively.
Our well depth for the 2$^2\Sigma^+$ and 3$^2\Sigma^+$ states are of 63.04\% and 90.30\% larger than those of Valance [50].

The large discrepancies between the Valence’s results and our values can be explained by two reason. First, Valence used the Hellmann-type model 
where the Hamiltonian is written in terms of one valence electron interacting with closed shell cores. Therefore, the core-core interaction potentials 
were neglected. This is certainly true for large interatomic distances but it is not accurate enough for small separations. 
Valence considered this shortcoming as not relevant to the charge exchange collision process they studied using the interaction potential   
where only intermediate and long-range distances are required. Although, without considering the core-core interaction one can get correct asymptotic limits, 
at intermediate distance this will affect the accuracy of the equilibrium distance and well depth. Second, the core-valence interactions were considered in our work using a formulation where, 
for each atom, the core polarization effects are described by an effective potential as described previously. The latter is modified by an $l$- dependent cut-off function $F$ to consider, in a different way 
the orbital symmetries. In the case of Valence,  the Hellmann pseudopotential is not "$l$" dependent and the parameters are optimized to obtain the two first experimental energy levels. 
Similarly, in the the absence of the core-core interaction, Valence considered that this lack of "$l$" dependence is not crucial for the charge exchange collision. In our approach,
in addition to the use of an $l$-dependent pseudoptential the parameters were optimized to reproduce with high accuracy the ionization potential and the experimental energy levels of many alkali excited states. 
Moreover, in our study, the interatomic separation grid is more extended as $R$ varies from 3 to 200 a.u. with a distance step of 0.1 a.u,   but Valence limited his calculation for internuclear distances ranging
from 5 a.u. to 25 a.u. It is important to note that the calculation time in our calculation is enormously reduced due to the use of the pseudopotential approach that leads these ionic molecules to one-electron systems. 
Calculations are performed on a desktop Pentium IV Acer computer where the time for one single distance is about 32 seconds. 

\section{Elastic collisions: Results and discussions}

Here we present results on  elastic collisions  of the discussed
ion-atom systems  by FORTRAN code \cite{press:nrcp} using  the well-known Numerov-Cooley algorithm \cite{cooley1961}.
This algorithm provides an efficient  technique for numerically solving a second order differential equation. It 
uses a three-point recursion relation to calculate first order derivative and makes use of exact second order derivative 
provided by the equation itself. To solve time-independent Schroedinger equation as a scattering problem, 
one has to propagate the Numerov-Cooley code from initially small internuclear separation $r$ to asymptotically large $r$ where 
the wave function corresponds to the state of a free particle and so behaves sinusoidally. The scattering $S$-matrix is then deduced 
by matching the asymptotic solution with the standard scattering boundary conditions. The correctness of the calculations is ensured 
by the unitarity of the $S$-matrix. The lower the energy larger is the asymptotic boundary. To initiate the propagation of the code 
from $r \simeq 0$ position, one has to set initially the values of the wave functions at two initial positions, then the code calculates 
the wave function for the third position, and then taking the the values of the wave functions for the  second and third position it calculates 
the wave function for the fourth position, and the process continues until the asymptotic boundary is reached. The initial boundary 
conditions are set by expanding the interaction potential for small $r$ and solving the Schroedinger equation analytically 
for $r \rightarrow 0$ limit. A good measure for the asymptotic scattering 
boundary (large $r$) can be found by setting the condition that  the  effective potential 
$V_{eff} = - C_6/r^6 + (\hbar^2/2\mu)  \ell (\ell + 1)/r^2$ for large $r$ becomes much less than 
the collision energy $E$ (at least one tenth of $E$).

We use the interaction potentials calculated
above as input data. In particular,  we focus on low energy collisions. Using 
well-known expansion of  continuum state in terms of partial waves,
the wave function $\psi_{\ell}$ ( k r) for $\ell$th  partial wave is
given by
\begin{equation} \left[ \frac{d^{2}}{dr^{2}} + k^{2} -
\frac{2\mu}{\hbar^{2}} V(r) - \frac{\ell (\ell +1)}{r^{2}} \right ]
\psi_{\ell} (kr) = 0
\end{equation}
subject to the standard scattering boundary condition
\begin{equation} \psi_{\ell}(kr)\sim \sin\left[ kr - \ell\pi/2 + \eta_{l}\right]
\end{equation}
Here $r$ denotes  the ion-atom separation, the wave number $k$ is
related to the collision energy $E$ by $E = \hbar^{2}$ $k^{2} /2 \mu$
and $\mu$ stands for the reduced mass of the ion-atom pair. The
total elastic scattering cross section is given by
\begin{equation}
\sigma_{el} = \frac{4\pi}{k^2}\sum_{\ell=0}^\infty
(2\ell+1)\sin^{2}(\eta_\ell).
\end{equation}

To know the relevant energy regimes where $s$- $p$- and $d$-wave
collisions are important, we have plotted in figure 2 the
centrifugal energies for first 3 partial waves ($\ell$ = 0, 1 and 2)
against ion-atom separations for ground state 1$^{2}\Sigma^{+}$ of
(NaCs)$^{+}$ and (LiCs)$^{+}$, respectively. For $d$-wave, the values
of centrifugal barrier are about 0.007 mK for (NaCs)$^{+}$ and about
0.06 mK for (LiCs)$^{+}$. These values  indicate   that the
potential energy barriers for low lying  higher partial waves are
very low for atom-ion systems allowing tunneling of the wave
function towards the inner region of the barriers. Unlike atom-atom
systems at low energy, a number of partial waves can significantly
contribute to the ion-atom scattering cross section at low energy.
We find that in order to get convergent results at milli- and micro-Kelvin regimes, 
the numerical calculation of scattering wave function needs to be extended to at least  10000 $a_0$ and $20000 a_0$, 
respectively. .  

In figure 3, the partial-wave cross sections for ground
1$^{2}\Sigma^{+}$ state collisions are plotted against collision
energy $E$  in K  for (NaCs)$^{+}$ and (LiCs)$^{+}$. As $ k \rightarrow
0$ , $s$-wave cross section becomes independent of energy while  $p$- and
$d$-wave cross sections vary in accordance with Wigner threshold laws.
According to Wigner threshold laws, as k $\rightarrow$ 0, the phase
shift for $\ell$th partial-wave  $\eta_l \sim k^{2\ell+1}$  if $\ell
\leq (n-3)/2$, otherwise $\eta_{l} \sim k^{n-2}$ for a long range
potential behaving as $1/r^{n}$. Since Born-Oppenheimer ion-atom
potential goes as $1/{r^4}$ in the asymptotic limit, as $k
\rightarrow 0$, $s$-wave scattering cross section becomes independent
of energy  while all other higher partial-wave cross sections
go as $\sim k^{2}$. As energy increases beyond the Wigner threshold law 
regime, the $s$-wave cross section exhibits  a minimum at a low energy which may be related to  Ramsauer-Townsend effect. 
It can be noticed from figure 3 that at Ramsauer minimum of $s$-wave scattering, $p$- and $d$-wave 
elastic cross sections are finite. This means that one can explore $p$- or $d$-wave interactions of ion-atom systems 
at the minimum position that occurs at a relatively low energy ($\mu$- and milli-Kelvin regime). 

In the Wigner threshold regime, hetero-nuclear
ion-atom collisions are dominated by elastic scattering processes
since the charge transfer reactions are highly suppressed in this
energy regime. Furthermore, resonant charge transfer collisions do
not arise in collision of an ion with an atom of different nucleus. 

It is worthwhile  to compare the elastic scattering results of hetero-nuclear
alkai ion-atom system to those of homo-nuclear alkali atom-ion system Na+Na$^{+}$ 
\cite{cote2000}. By comparing our results in figure 3  with the figures 2 and 3 of Ref. 
\cite{cote2000}, we notice that elastic scattering cross sections of  both homo- and hetero-nuclear ion-atom systems 
at ultralow energies are comparable and primarily governed by Winger threshold laws. However, 
unlike hetero-nuclear systems, the resonant charge transfer cross section in homo-nuclear Na+Na$^{+}$ is also of the
same order as that of elastic one at ultra-low energies.  Resonant charge exchange cold collisions between Yb and Yb$^{+}$ 
are experimentally observed and found to be a dominant process in this homo-nuclear system \cite{prl:2009:vuletic}.

In figure 4, we have plotted logarithm of total elastic scattering
cross sections ($\sigma_{tot}$) expressed in a.u. as a function of
logarithm of $E$ in K for ground state collisions between Li and
Cs$^{+}$ and between Na and Cs$^{+}$, respectively. For both
(LiCs)$^{+}$ and (NaCs)$^{+}$, to calculate $\sigma_{tot}$, we
require more than 60 partial waves to get converging results for
energies greater than 1 mK. This is because, as the collision energy increases, more and more partial waves start to contribute to
$\sigma_{tot}$. In order for a particular partial wave $\ell$ to contribute appreciably at a given energy $E$,
the centrifugal barrier height corresponding to this partial wave should not be too high
compared to the energy. An estimate of the minimum number of partial waves required for obtaining convergent results may be made by 
taking the height of the centrifugal barrier to be higher than the energy
by one order of magnitude. In this context, it is worth mentioning that the centrifugal barrier height for a partial wave
of an ion-atom system is usually much lower than the corresponding height of atom-atom system, therefore larger
number of partial waves contribute to $\sigma_{tot}$ in case of an ion-atom system.
In the large energy limit, the cross
section behaves as
\begin{equation} \sigma_{tot} \sim \pi
\left( \frac{\mu C_{4}^{2}}{\hbar^{2}}\right)^\frac{1}{3} \left(1 +
\frac{\pi^{2}}{16}\right) E^{-\frac{1}{3}}\label{sig}
\end{equation}

Thus, in the large energy regime the slope of the  logarithm of
$\sigma_{tot}$ as a function of $\log E$ is a straight line obeying
the equation $\sigma_{tot}(E) = - (1/3) E + c_{E}$ where the slope
of the line is $- 1/3$ and the intercept of the line on the energy
axis is solely determined by the $C_{4}$ coefficient of long-range
part of the potential, or equivalently by the characteristic length
scale $\beta$ of the potentials or the polarizability of the neutral
atom interacting with the ion. As shown in figure 4, we have
numerically verified this E$^{-1/3}$ law for both Li-Cs$^{+}$ and
Na-Cs$^{+}$ systems. The linear fit to the the plot of numerically
calculated $\sigma_{tot}$ against in the large energy regime shows
that the value of the slope is quite close to the actual value of
1/3.
 Since the dipole polarizability of Li and Na is not much different, one can
expect that for both Li-Cs$^{+}$ and Na-Cs$^{+}$ systems, the
energy-dependence of $\sigma_{tot}$ at large energy should be
similar. In fact, figure 4 shows that for energies much greater 1
$\mu$K, both systems exhibit similar asymptotic energy dependence.

\begin{figure}
        \includegraphics[width=3.65in]{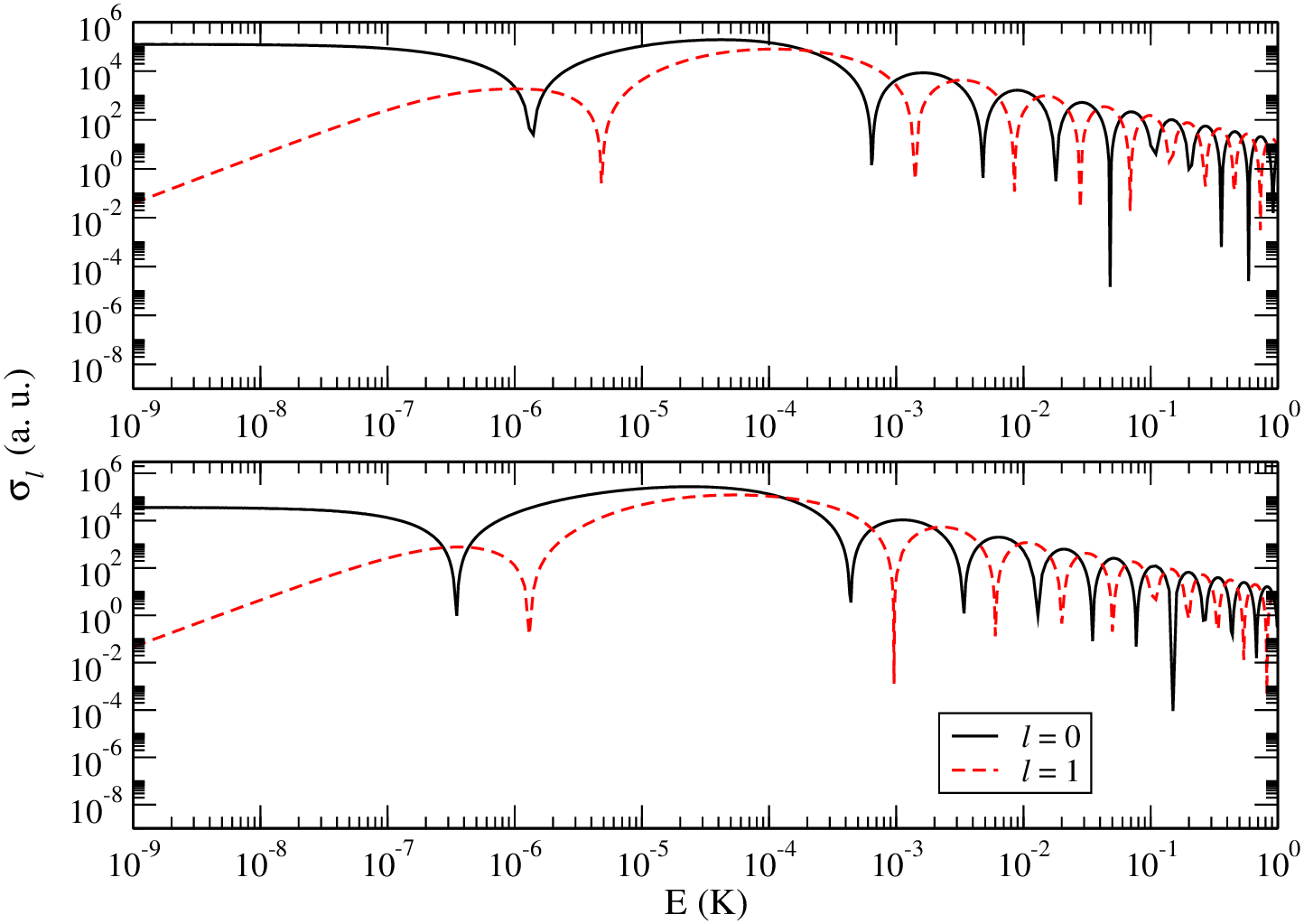}
        \caption{(Color online)Partial wave cross sections for $^{6}$Li + Rb$^{+}$ (1$^{2}\Sigma^{+}$) and $^{7}$Li + Rb$^{+}$
        (1$^{2}\Sigma^{+}$) collision are plotted as a function of E (in K) for $\ell$ = 0 (solid) and $\ell$ = 1 (dashed) in
        upper and lower panel, respectively.}
\end{figure}

Finally, we consider  the collision of $^{85}$Rb$^{+}$ with the two
isotopes of Li atom, namely, $^{6}$Li and $^{7}$Li, and the results
are shown in figures 5 and 6.  The purpose here is to investigate
the isotopic effects of Li on low energy collisions with
$^{85}$Rb$^{+}$ ion . In figure 5 we have plotted logarithm (to the
base of 10) of $\sigma_{tot}$ as a function of logarithm (to the
base of 10) of E. Results clearly show that the patterns are same
for $^{6}$Li and $^{7}$Li colliding with $^{85}$Rb$^{+}$, but they
differ in magnitudes, particularly in the low energy regime. Figure
6 exhibits the variation of $s$- and $p$-wave scattering cross sections
against energy in log-log scale for  both isotopes of Li colliding
with Rb$^{+}$. Again, the results are qualitatively similar for both
the isotopes, but they differ slightly quantitatively. Comparing the
upper and lower panels in figure 6, one can notice that the Wigner
threshold law regime for $^{6}$Li can be attained at slightly higher
energy than that for $^{7}$Li.

\section{Conclusions}

In conclusion, we have studied  elastic scattering between alkali ion and alkali atom of different nuclei
of three  ion-atom systems, namely, Li + Cs$^{+}$, Na +
Cs$^{+}$, and Li + Rb$^{+}$. We have calculated the interaction potentials and spectroscopic constants 
of these systems. We have presented a detailed study of elastic collision physics over a wide range of energies, showing the 
onset of Wigner threshold regime at ultra low energy and the 1/3 law of scattering at higher energy regime. 
The low energy scattering results presented here
may be useful for future exploration for radiative- or photo-associative formation
of cold diatomic molecular ions from these three  ion-atom systems.  The
colliding ground state atom-ion pair corresponding to the continuum
of 1$^{2}\Sigma^{+}$ may be photoassociated to form a bound state in
second excited potential 3$^{2}\Sigma^{+}$ by using a laser of
appropriate frequency. This is promising because the transition is
dipole allowed and is similar to atomic transition. The formation of
such cold molecular ions by radiative processes will pave new
directions to ultracold chemistry.

\vspace{0.5cm}

\noindent 
{\bf Acknowledgment} \\
This work is jointly supported by Department of Science and Technology (DST), Ministry of Science and Technology, Government of India and Ministry of 
Higher Education and Scientific Research (MHESR), Government of Tunisia, under an India-Tunisia project for bilateral sientific cooperation.


\section*{References}

\begin{thebibliography}{10}

\bibitem{prl:2009:vuletic} Grier A T, Certina M, Orucevic F and Vuletic V 2009  {\it Phys. Rev. Lett.} 
{\bf 102}, 223201

\bibitem{aymar:prl:2011} Hall F H J, Aymar M, Bouloufa N, Dulieu O and Willitsch S 2011 
{\it Phys. Rev. Lett.} {\bf 107} 243202

\bibitem{molphys} Hall F H J, Eberle P, Hegi G, Raoult M, Aymar M, Dulieu O and Willitsch S 2013  
{\it Mol. Phys.} {\bf 111}, 2020

\bibitem{hudson:prl:2011} Rellergert W, Sullivan S, Kotochigova S, Petrov A, Chen K, Schowalter S and Hudson E 2011 {\it Phys. Rev. Lett.} {\bf 107} 243201

\bibitem{hudson:prl:2012} Sullivan S, Rellergert W, Kotochigova S and Hudson E 2012 {\it Phys.
Rev. Lett.} {\bf 109} 223002 

\bibitem{Zipkes1} Zipkes C, Palzer S, Sias C and Kohll M 2010 {\it Nature} {\bf 464} 388

\bibitem{Zipkes2} Zipkes C, Palzer S, Ratschbacher L, Sias C and Kohll M 2010 {\it Phys. Rev. Lett.}
{\bf 105}, 133201

\bibitem{Zipkes3} Ratschbacher L, Zipkes C, Sias C and Kohll M 2013 {\it Nature Phys.} {\bf 8} 649

\bibitem{Mukaiyama:pra:2013} Haze S, Hata A, Fujinaga M and Mukaiyama T 2013 {\it Phys. Rev. A}  {\bf 87} 052715

\bibitem{schmid:2012:rsi} Schmid S, Härter A, Frisch A, Hoinka S and Hecker-Denschlag J 2012 
 {\it Rev. Sci. Instrum.} {\bf 83} 053108

\bibitem{Denschlag:prl:2012} Haerter A, Kruekow A, Brunner A, Schnitzler W, Schmid S and Denschlag J H 2012  {\it Phys. Rev. Lett.} {\bf 109} 123201

\bibitem{Narducci:pra:2012} Sivarajah I, Goodman D S, Wells J E, Narducci F A and Smith W W 2012 {\it Phys. Rev. A} {\bf 86}  063419

\bibitem{schmid}  Schmid S, H\"arter A and Denschlag J H 2010 {\it Phys. Rev. Lett.} {\bf 105} 133202

\bibitem{rangwala:apl:2012} Ravi K, Lee S, Sharma A, Werth G and Rangwala S 2012  
 {\it Appl. Phys. B} {\bf 107} 971 

\bibitem{rangwala:natcom:2012} Ravi K, Lee S, Sharma A, Werth G and Rangwala S 2012  
{\it Nature Comm.} {\bf 3}

\bibitem{rangwala:pra:2013} Lee S, Ravi K and Rangwala S 2013 {\it Phys. Rev. A}  {\bf 87} 052701

\bibitem{hytrap} Smith W W, Marakov O P and Lin J 2005 {\it J. Mod. Opts.} {\bf 52} 2253

\bibitem{prl111:2013:schmidt-kaler} Bissbort U, Cocks D, Negretti A, Idziaszek Z, Calarco T, Schmidt-Kaler F, Hofstetter W and Gerritsma R 2013 
{\it Phys. Rev. Lett.} {\bf 111} 080501

\bibitem{prl109:2012:schmidt-kaler} Gerritsma R, Negretti A, Doerk H, Idziaszek Z, Calarco T and Schmidt-Kaler F 2012 
{\it Phys. Rev. Lett.} {\bf 109}  080402 

\bibitem{pra89:2014:negretti} Joger J, Negretti A and Gerritsma R 2014 {\it Phys. Rev. A} {\bf 89} 063621 

\bibitem{pra81:2010:busch} J. Goold J, H. Doerk H, Z. Idziaszek Z, T. Calarco T and T. Busch T 2010 {\it Phys. Rev. A} {\bf 81} 041601

\bibitem{cote2000} C\^ot\'e R 2000 {\it Phys. Rev. Lett.} {\bf 85} 5316

\bibitem{njp17:2015:aymar}  Silva Jr H da, Raoult M, Aymar M and Dulieu O 2015 {\it New J. Phys} {\bf 17}

\bibitem{bound} C\^ot\'e R, Kharchenko V and M. D. Lukin M D 2002 {\it Phys. Rev. Lett.} {\bf 89} 093001

\bibitem{Rakshit2011} Rakshit A and Deb B 2011 {\it Phys. Rev. A} {\bf 83} 022703

\bibitem{jcp:2011:dulieu} Aymar M, Guerout R and Dulieu O 2011 {\it J. Chem. Phys.} {\bf 135} 064305

\bibitem{arXiv:1409.1192} Tomza M, Koch C P and Moszynski R 2015  {\it arXiv:1409.1192}

\bibitem{pra91goodman} D. S. Goodman D S, Wells J E, Kwolek J M, Blumel R, Narducci F A and Smith W W 2015 {\it Phys. Rev. A} {\bf 91} 012709

\bibitem{njp} Bodo E, Zhang P and Dalgarno A 2008 {\it New J. Phys.} {\bf 10} 033024

\bibitem{Zhang2009} Zhang P, Bodo E and Dalgarno A 2009 {\it J. Phys. Chem. A} {\bf 113} 15085

\bibitem{Zygelman1989} Zygelman B, Dalgarno A, Kimura M and Lane N F 1989 {\it Phys. Rev. A} {\bf 40} 2340

\bibitem{pra79} Idziaszek Z, Calarco T, Julienne P S and Simoni A 2009 {\it Phys. Rev. A} {\bf 79} 010702(R)

\bibitem{bgao} Gao  2010 {\it Phys. Rev. Lett.} {\bf 104} 213201

\bibitem{bgao2012} Li M and Gao B 2012 {\it Phys. Rev. A} {\bf 86} 012707 

\bibitem{pra81:2010:calarco} Doerk-Bendig H, Idziaszek Z and Calarco T 2010  {\it Phys. Rev. A} {\bf 81} 012708

\bibitem{contphys} Harter A and Denschlag J H 2014 {\it Contemporary Phys.} {\bf 55} 33

\bibitem{pccp13:2011:belyaev} Tacconi M, Gianturco F A and Belyaev A K 2011 {\it Phys. Chem. Chem. Phys.} {\bf 13} 19156 

\bibitem{pra87:2013:belyaev}  Sayfutyarova E R, Buchachenko A A, Yakovleva S A and Belyaev A K 2013 {\it Phys. Rev. A} {\bf 87} 052717

\bibitem{jpb47:2014:McCann} McLaughlin B M, Lamb H D L, Lane T C and McCann J F 2014 
{\it J. Phys.  B: At. Mol. Opt. Phys.} {\bf 47} 145201 
 
\bibitem{jpb47:2014:zygelman} Zygelman B, Zelimir L and Hudson E R 2014 {\it J. Phys. B: At., Molec. Opt. Phys.} {\bf 47} 015301 

\bibitem{pra86:2012:McCann} Lamb H, McCann J, McLaughlin B, Goold J, Wells N and Lane I 2012 {\it Phys. Rev. A} {\bf 86} 022716 

\bibitem{naturecomm} Huber T, Lambrecht A, Schmidt J, Karpa L and Schaetz T 2014  {\it Nature} {\bf 5} 5587

\bibitem{cooley1961} J. W. Cooley 1961 {\it Math. Comp.} {\bf 15}, 363 

\bibitem{Barthelat1975} Barthelat J C and Durand Ph 1975 {\it Theor Chem. Acta} {\bf 38} 283

\bibitem{Ghanmi1} Ghanmi C, Farjallah M and Berriche H 2012 {\it J. Mol. Spect.} {\bf 235} 158 

\bibitem{Ghanmi2} Ghanmi C, Bouzouita H, Berriche H and Ben Ouada H 2006 {\it J. Mol. STruct. THEOCHEM} {\bf 777} 81

\bibitem{Ghanmi2012} Ghanmi C, Farjallah M and Berriche H 2012, {\it Int. J. Quant. Chem.} {\bf 112} 2403

\bibitem{Berriche2003} Berriche H 2003 {\it J. Mol. Struct. THEOCHEM} {\bf 663} 101

\bibitem{Berriche2005} Berriche H, Ghanmi C and Ben Ouada H 2005 {\it J. Mol. Spect.} {\bf 230} 161

\bibitem{Valance1978} Valance A 1978 {\it J. Chem. Phys.} {\bf 69} 355

\bibitem{Muller1984} Muller W, Flesh J and Meyer W 1984 {\it J. Chem. Phys.} {\bf 80} 3297

\bibitem{Fouc1992} Foucrault M, Millie Ph and Daudey J P 1992  {\it J. Chem. Phys.} {\bf 96} 1257

\bibitem{Pavolini1989} Pavolini D, Gustavsson T, Spiegelmann F and Daudey J P 1989 {\it J. Phys. B At. Mol. Opt. Phys.} {\bf 22} 1721

\bibitem{Bellomonte1974} Bellomonte L, Cavaliere P and Ferrante G 1974 {\it J. Chem. Phys.} {\bf 61} 3225

\bibitem{Carlson1980} Carlson N W, Taylor A J and Shawlow A L 1980 {\it Phys. Rev. Lett.} {\bf 45} 18

\bibitem{Patil2000} Patil S H and Tang K T 2000 {\it J. Chem. Phys.} {\bf 113} 676

\bibitem{Azizi1} Azizi S, Aymar M and Dulieu O 2007 {\it AIP. Conf. Proc.} {\bf 935} 164

\bibitem{Khelifi2011} Khelifi N, Dardouri R and Al-Dossary O M 2011 {\it J. Appl. Spect.} {\bf 78} 11

\bibitem{press:nrcp} Press W H, Teukolsky S A, Vetterling W T, Flannery B P 2007  
{\it Numerical Recipes in FORTRAN: The Art of Scientific Computing}  {\bf 3rd edition} Cambridge University Press 


\end{thebibliography}
\end{document}